\documentclass[12pt,preprint]{aastex}

\newcommand{\degg}{\hbox{$^\circ$}}

\newcommand{\arcs}{\hbox{$^{\prime\prime}$}}

\newcommand{\et}{et al.\ }
\newcommand{\xte}{{\it RXTE}}

\newcommand{\xmm}{{\it XMM-Newton}}

\newcommand{\sax}{{\it BeppoSAX}}

\newcommand{\asca}{{\it ASCA}}

\newcommand{\ls}
{\mathrel{\hbox{\rlap{\hbox{\lower4pt\hbox{$\sim$}}}\hbox{$<$}}}}
\newcommand{\gs}
{\mathrel{\hbox{\rlap{\hbox{\lower4pt\hbox{$\sim$}}}\hbox{$>$}}}}

\begin{document}

\title{A Massive X-ray Outflow from the Quasar PDS 456}
\shorttitle{A Massive X-ray Outflow in PDS 456}
\shortauthors{Reeves \et}
\author{ J.N. Reeves\altaffilmark{1,2,3}, P.T. O'Brien\altaffilmark{3}, 
M.J. Ward\altaffilmark{3}}
\email{jnr@milkyway.gsfc.nasa.gov}

\altaffiltext{1}{Laboratory for High Energy Astrophysics, Code 662, 
NASA Goddard Space Flight Center, Greenbelt Road, Greenbelt, MD 20771, USA.}

\altaffiltext{2}{Universities Space Research Association}

\altaffiltext{3}{X-ray and Observational Astronomy Group; University of
Leicester; Leicester LE1 7RH; United Kingdom}

\begin{abstract}

We report on XMM-Newton spectroscopic observations of the luminous,
radio-quiet quasar PDS 456. The hard X-ray spectrum of PDS 456
shows a deep absorption trough (constituting 50\% of the continuum)
at energies above 7 keV in the quasar rest frame, which can be attributed 
to a series of blue-shifted K-shell absorption edges due 
to highly ionized iron. The higher resolution soft X-ray  
RGS spectrum exhibits a broad absorption line feature near 1 keV, 
which can be modeled by a blend of
L-shell transitions from highly ionized iron (Fe \textsc{xvii} --
\textsc{xxiv}).
 An extreme outflow velocity of $\sim50000$~km~s$^{-1}$ is required to
model the K and L shell iron absorption present in the XMM-Newton data. 
Overall, a large column density ($N_{H}=5\times10^{23}$cm$^{-2}$) of
highly ionized gas (log~$\xi=2.5$) is required in PDS 456. 
A high mass outflow rate 
of $\sim10~M_{\odot}$~year$^{-1}$ (assuming a conservative 
outflow covering factor of 0.1 steradian) is derived, which is 
of the same order as the overall mass accretion rate in PDS 456.
The kinetic energy of the outflow 
represents a substantial fraction ($\sim10$\%) of the quasar 
energy budget, whilst the large column and 
outflow velocity place PDS 456 towards the extreme end of the 
broad absorption line quasar population. 

\end{abstract}


\section{Introduction}

PDS 456 is a luminous, low redshift ($z=0.184$) radio-quiet quasar
identified in 1997 \citep{Tor97}. The optical and
infra-red spectra \citep{Simpson99}
show broad Balmer and Paschen lines (e.g. H$\beta$ FWHM 3000 km~s$^{-1}$),
strong Fe \textsc{ii}, a hard (de-reddened)
optical continuum ($f_{\nu} \propto \nu^{-0.1\pm0.1}$), and one of the
strongest `big blue bumps' of any AGN \citep{Simpson99,R00}. 
PDS~456 has a de-reddened, absolute blue magnitude of
M$_{B}\approx -27$ \citep{Simpson99},
making it at least as luminous as the radio-loud quasar 3C~273.
Indeed PDS 456 is the most luminous known AGN in the local Universe
(z\ $<0.3$), its luminosity being more typical of quasars at z=2-3, at the
peak of the quasar luminosity function. 

PDS 456 was first detected as the X-ray source
RXS J172819.3-141600 in the {\it ROSAT} All Sky Survey \citep{Voges99}. 
Subsequent \asca\ and \xte\ and \sax\ observations 
showed that it was highly X-ray variable \citep{R00, R02},
requiring that the X-ray source was very compact
($<3$~Schwarzschild radii) and accreting with a unusually high
efficiency (between 6\% and 41\%), close to the limit for rotating Kerr 
black hole \citep{Thorne74}. 
The X-ray spectrum of PDS 456, obtained by \asca\ and \xte\, also
appeared to be unusual for a quasar. A strong edge like feature 
was present in the iron K-shell band, either due to a highly
ionized absorber or reflector, whilst absorption towards the
soft X-ray band was also observed \citep{R00}. 
An \xmm\ observation of PDS 456 was subsequently obtained in AO-1, 
in order to obtain high signal to noise spectra of both the iron K-shell
complex with the EPIC CCD detectors and a high resolution spectrum of the
soft X-ray absorber with the \xmm\ RGS (Reflection Grating Spectrometer). 
Here we report on the X-ray spectra obtained through the 
\xmm\ observations of PDS 456, 
which reveal evidence for a massive outflow, with 
a velocity of $\sim50000$~km~s$^{-1}$. 
A previous paper \citep{R02} discussed in detail the 
X-ray variability during the \xmm\ observation and a simultaneous 
week long \sax\ observation.


\section{The X-ray Observations}

PDS 456 was observed by \xmm\ on 2001 February 26th 
with a duration of 40 ks. Data were taken with the EPIC-pn detector 
\citep{Struder01} in Full Window Mode and with the EPIC-MOS detector 
\citep{Turner01} in Large Window mode.  The data was
reduced using version 5.4 of the XMM-SAS software. Note that the 
time-averaged, 0.5-10 keV flux of PDS~456 during the observation was 
$9.7\times10^{-12}$~ergs~cm$^{-2}$~s$^{-1}$; at this flux level the 
effect of photon pile-up is small ($<1$\%).
All time intervals were included, as the observation 
contained no background flares. Data were selected using event patterns 0-12 
(for the MOS) and pattern 0-4 (for the pn) and only good X-ray events 
were included. The spectra were extracted from circular
source regions of 40\arcs\ radius, whilst background spectra were
extracted from an offset circle, close to PDS 456, but free of any
background sources. Response matrices and ancillary response
files were generated using the SAS tasks \textsc{rmfgen} and 
\textsc{arfgen} respectively.


Background subtracted spectra were fitted using \textsc{xspec} v11.2,
including data over the energy ranges 0.4 to 12 keV for the
EPIC-pn and 0.4 keV to 10 keV for the co-added EPIC-MOS spectrum. A
Galactic absorption column of $N_{H}=2\times10^{21}$~cm$^{-2}$ 
\citep{Dickey90} was included in all the fits. 
A single absorbed power-law fit 
(photon index, $\Gamma=2.0$) to both MOS and PN data is 
clearly inadequate (reduced $\chi^{2}=10$). Figure 1 shows 
the data/model ratio to this power-law, a strong 
soft excess is observed below 1 keV and a
large drop in counts is observed above 7 keV in the iron K-shell band. 
The MOS and pn spectra are consistent, 
except below 0.5~keV where there is a small
divergence in spectral slope.

\section{The Ionized Absorber in PDS~456}

Initially, to model the iron K-shell band,
the data were fitted from 2-12 keV (or 2-10 keV for
EPIC-MOS), so that the effects of the large soft excess are negligible. 
As the EPIC-pn data contained significantly more counts than the EPIC-MOS
at the highest energies, 
we proceeded to fit the EPIC-pn data only.
The ratio spectrum above 2 keV to a power-law fitted over the 2-5 keV range 
(i.e. outside of the iron K band) is shown in the inset to Figure 1.
A strong deficit of counts occurs above 7 keV, 
where one would expect to observe
K-shell absorption edges from highly ionized iron. 
Indeed a simple Galactic absorbed power-law produced a poor fit 
($\chi^{2}/dof=1001/746$, null hypothesis probability $9\times10^{-10}$). 

Adding three absorption edges between 7 and 10 keV significantly 
improves the fit ($\chi^{2}/dof=739/740$), modeling the broad Fe 
absorption feature present in the X-ray spectrum. 
The energies (in the quasar rest frame) 
and optical depths of the edges are then;
$E=9.3\pm0.1$~keV, $E=8.5\pm0.2$~keV and $E=7.6\pm0.3$~keV with 
$\tau=0.56\pm0.11$, $\tau=0.26\pm0.11$, $\tau=0.24\pm0.08$
respectively. The breadth of the absorption feature suggests that a wide range 
of ionizations may be present, whilst the depth of the feature
implies a column of gas of 
$N_{H} \sim 10^{24}$~cm$^{-2}$. Note that 
the upper limit to a narrow (i.e. unresolved) Fe emission line (in the energy 
range 6.4 - 6.97 keV) is $<$~12~eV, implying that the solid angle subtended
by the absorber must be small ($\Omega<0.5$~steradian). Note that an 
extreme broad iron K$\alpha$ emission line from an accretion disk 
cannot model the shape of the 
spectrum from 2-12 keV. If one attempts to model the spectrum with a 
Fe line profile from a rapidly rotating Kerr black hole \citep{Laor91}, the 
physical parameters required are unphysical (disk emissivity $R^{-10}$, 
inclination=90\degg, equivalent width $>1$~keV), because the line is 
(unsuccessfully) attempting to model the broad absorption dip above 7 keV with 
the blue-wing of the line profile.  

To obtain a more physical representation of the Fe K-shell absorber, we 
used a grid of models generated by the \textsc{xstar} photoionization code 
\citep{Kallman96} to fit the EPIC-pn data above 2 keV. 
Solar elemental abundances 
were assumed, whilst a turbulence velocity of 1000~km~s$^{-1}$ was used. A one 
zone photoionization model was adopted, with the column density ($N_{\rm H}$), 
ionization parameter ($\xi$), partial covering fraction ($f$) and 
outflow velocity ($v_{\rm out}$) as free parameters. 
An effective hydrogen column density (for a solar abundance H/Fe ratio)
of $5.7^{+2.0}_{-2.5}\times10^{23}$~cm$^{-2}$ is required, 
together with an ionization parameter of ${\rm log}~\xi=2.5\pm0.3$ 
and a partial 
covering fraction of $f\sim0.6$, whilst the power-law continuum steepens 
to $\Gamma=2.40\pm0.05$. At this ionization state most of the 
absorption is due to Fe~\textsc{xvii} - Fe~\textsc{xxiv} 
(i.e. where the L-shell is partially filled). 
The best fit spectrum modeled by \textsc{xstar} 
is shown in Figure 2 (panel a), 
the depth and energy of the absorption feature 
being well matched, whilst the partial covering warm absorber also 
models the convex spectral curvature (through the increase in opacity 
at lower energies due to iron L-shell absorption)
present in the EPIC-pn data between 2-6 keV. Overall, the fit statistic 
for this model is excellent ($\chi^{2}/dof=717/740$).

Importantly a large outflow velocity 
of 47000~km~s$^{-1}$ (90\% confidence limits, 36000~km~s$^{-1}$ -- 
82000~km~s$^{-1}$) is needed to model the spectrum. Although the exact 
outflow velocity is not well constrained by the EPIC-pn data 
alone, a significant outflow velocity is still required. Figure 2(b) 
illustrates this point, if one fixes the absorber velocity to 
zero, then the model fit becomes poor as the iron K 
edges then lie at a significantly lower energy than is observed 
in the EPIC data. Indeed if one tries to increase the energy of the 
model absorption feature by 
increasing the ionization state of the absorber (so that He and H-like iron 
dominates), the result is a worse fit still, partly because the 
absorption edge feature becomes weaker (as a significant fraction of iron 
becomes fully ionized) and also because the 
most prominent absorption features are now the strong K$\alpha$ and K$\beta$ 
absorption lines due to Fe \textsc{xxvi} 
which are not present in the actual \xmm\ data 
(Figure 2c). Thus a significant outflow velocity is {\it required} by the 
data, indeed one can place a conservative lower-limit 
(at 99.9\% confidence) on the velocity of 24000~km~s$^{-1}$ (or $\sim0.08c$). 
 
\subsection{The soft X-ray RGS Spectrum of PDS 456}

Extrapolation of the hard X-ray spectrum to lower energies (down to 
0.4 keV), leaves clear residuals in the EPIC MOS and pn data (e.g. Figure 1).
A strong soft excess is present below 1 keV, whilst residuals due to 
an additional absorber may also be present in the soft X-ray band. 
To investigate this at higher spectral resolution, we used data 
available from the Reflection Grating Spectrometer (RGS) 
below 2 keV. The most noticeable feature present is a broad
absorption trough observed between 12-15~\AA\ 
(see Figure 3). The feature 
is highly significant, being detected at $>99.99\%$ confidence. When 
parameterized by a simple broad Gaussian shaped absorption line, the 
best fit parameters (in the quasar rest frame) are $E=1.07\pm0.02$~keV with an 
equivalent width of $26\pm10$~eV and a velocity width of 
$9000\pm3000$~km~s$^{-1}$.

As in the previous section, we used our grid of photoionization models 
to characterize the properties of the absorbing 
gas, apparent in the RGS spectrum. A lower column 
density ($N_{\rm H}\sim5\times10^{22}$~cm$^{-2}$)  
is required to fit the data, whilst both a similar 
ionization parameter (${\rm log}~\xi=2.6\pm0.5$) and outflow velocity 
(with $v_{\rm out}=57000^{+8000}_{-10000}$~km~s$^{-1}$) are needed, as per 
the fits to the EPIC-pn data. 
Partial covering is not required in the RGS fits. 
The photon index obtained ($\Gamma=2.4$, after correcting for absorption) 
is consistent with the EPIC-pn data, whilst a 
soft blackbody component (with $kT_{BB}=80\pm6$~eV) is also needed in the 
fit. The lower column density derived from the RGS fit indicates that 
the absorber is patchy and thus is likely to be located 
quite close to the X-ray source. Thus we see a higher 
(partially covered) column along the line of sight towards the 
hard X-ray continuum, whilst the column towards the soft X-ray continuum 
is lower. 

Inspection of the best-fit photoionization model implies that a blend of 
L-shell transitions from Fe~\textsc{xvii} -- Fe~\textsc{xxiv} 
as well as K-shell absorption from Ne~\textsc{ix/x} is 
responsible for the broad absorption feature at 12-15~\AA. 
A weak Ly-$\alpha$ absorption line due to O~\textsc{viii} 
may also be present at 19~\AA.
The outflow velocity derived appears robust, due to the 
positioning of the broad absorption feature. If one fixes the absorber 
velocity at zero in the quasar rest-frame, 
then the Fe L-shell absorption {\it in the model} is located at energies below 
the observed absorption trough, whilst increasing the ionization of the 
absorber only serves to significantly weaken the soft X-ray absorber. 

\section{Discussion}

X-ray spectroscopy of PDS 456 with \xmm\ has revealed a high column density 
($N_{\rm H}=5\times10^{23}$~cm$^{-2}$), highly ionized (log~$\xi=2.5$) 
warm absorber, its main signature being 
a series of K and L shell absorption edges and lines due to iron 
in the ionization range Fe~\textsc{xvii} -- Fe~\textsc{xxiv}. Crucially, 
a near relativistic outflow 
velocity is required, of the order 50000~km~s$^{-1}$. 
This velocity is similar to the actual systemic velocity of PDS 456 
(at $z=0.184$), therefore it is possible that the absorbing material is 
unrelated to the quasar, and is local to our Galaxy.
However we consider it extremely 
unlikely that such a high column, highly ionized absorber 
could occur by chance along the local line of sight.
Furthermore, although the Galactic column towards PDS 456 is 
moderately high ($N_{\rm H}=2\times10^{21}$cm$^{-2}$), 
this {\it neutral} absorption does not contribute 
towards the features seen in the iron K and L-shell bands, and in any 
event has been accounted for in all our modeling. 
Therefore it is highly 
probable that the absorber is intrinsic to the quasar PDS~456.

\subsection{A Massive Outflow in PDS 456}

The combination of a large column density, high ionization 
and a fast outflow velocity in PDS 456 indicates that the absorber is 
located close to the quasar's central engine and that its mass outflow rate 
is significant relative the to the accretion rate.
A {\it maximum} distance to the absorber can be calculated 
on the condition that its thickness 
($\Delta R$) cannot exceed its distance from the ionizing source ($R$), 
i.e. $\Delta R/R<1$. This 
requires that $R<10^{4}R_{g}$ (where $R_{g}=GM/c^{2}$) for a black hole mass 
of $10^{9}M_{\odot}$ for PDS 456 \citep{R00}, 
thus it is located within the BLR. 
One likely possibility is that the absorber arises 
through a disk driven outflow or wind. 
The exact origin of these winds and the mechanism responsible for driving them 
(radiation and/or magnetic) is unclear. For 
Broad Absorption Line (BAL) QSOs, line driven winds 
have been proposed \citep{Murray95, Proga00} 
as well as centrifugal driving along open magnetic field lines 
\citep{Bott97}.

Using conservation of mass, the outflow rate ($\dot{M}_{\rm out}$) for an 
outflow subtending a solid angle of
$\Omega$~steradian with a constant velocity $v_{\rm out}$ is given by:-

$\dot{M}_{\rm out}=\Omega n r^{2} v_{\rm out} m_{\rm p}$

where $n$ is the density, $r$ the distance from the X-ray source and 
$m_{\rm p}$ the proton mass. Now the ionization parameter $\xi$ is defined as 
$\xi=L_{\rm X}/nr^{2}$, so we can obtain:-

$\dot{M}_{\rm out}=\Omega v_{\rm out} m_{\rm p} L_{\rm X}/\xi$

From the observations, the hard X-ray luminosity of PDS 456 (above 2 keV) is 
$\sim10^{45}$~erg~s$^{-1}$, the ionization parameter is $\xi\sim1000$, 
whilst $v=50000$~km~s$^{-1}$.  
From the lack of any iron K line emission  
it is likely that the outflow subtends a solid angle of 
$<0.5$~steradian. However even for a conservative value of 
0.1~steradian, the mass outflow rate derived is 
$\dot{M}_{out}=8\times10^{26}$~g~s$^{-1}$ or $\sim10M_{\odot}$~year$^{-1}$. 
A lower limit of $\sim5M_{\odot}$~year$^{-1}$ is 
obtained using the 99.9\% confidence lower bound on the outflow
velocity of $v_{\rm out}=24000$~km~s$^{-1}$. 
Note the outflow rate calculated 
assumes the gas is in equlibrium, 
if the gas is being accelerated then the actual rate may differ.

PDS 456 is an extremely luminous quasar with a bolometric luminosity of 
$L_{bol}\sim10^{47}$~erg~s$^{-1}$ and a black hole mass of
$10^{9}{\rm M}_{\odot}$ assuming accretion occurs near the 
Eddington limit \citep{Simpson99, R00}. The mass accretion rate 
for PDS 456, given by $\dot{M}=L_{bol}/\eta c^{2}$, is  
$\sim20M_{\odot}$~year$^{-1}$, for an efficiency of $\eta=0.1$.  
Thus it seems likely the outflow rate is of the same order as the mass 
accretion rate in PDS 456.
Similarly the kinetic output of 
the outflow in PDS 456 is large, about $10^{46}$~erg~s$^{-1}$ 
corresponding to 10\% of the bolometric output of the quasar. 
The high velocity also implies that the outflow is likely to 
be driven from the accretion disk at a fairly small radius. For a 
Keplerian disk, the launch radius for the outflow is simply 
$R_{\rm out}=c^{2}/v_{\rm esc}^{2} R_{g}$, 
where $v_{\rm esc}$ is the escape velocity of the material.
Assuming that $v_{\rm esc} \sim v_{\rm out}$
(although some acceleration of the matter will 
probably occur), this implies that the outflow in PDS 456 is launched from a 
radius of $\sim40R_{g}$ (about $10^{16}$~cm for a $10^{9}$M$_{\odot}$ 
black hole). Indeed the launch radius may be smaller if the gas 
is not in simple equilibrium.

PDS 456 appears to be in a small but growing group of AGN to show 
extreme absorption properties in the iron K-shell band. The lensed 
BAL quasar APM~08279+5255 has been found to exhibit iron K 
absorption lines and/or edges \citep{Chartas02, Hasinger02} 
in both {\it Chandra} and \xmm\ observations. 
A similar absorber has also been reported 
in the BAL quasar PG 1115+080 \citep{Chartas03}. 
The outflow velocities are similar to PDS 456, in the range $0.1c-0.4c$. 
Given this apparent association between high velocity 
X-ray outflowing gas and the UV BALs, one 
may predict to see BAL features in the UV spectrum of PDS 456. Indeed a 
{\it HST-STIS} spectrum of PDS 456 (O'Brien \et 2003, in preparation) 
reveals that there may 
be cooler gas in the outflow; a BAL is seen in Lyman-$\alpha$ with 
an outflow velocity of 12000 -- 22000~km~s$^{-1}$, whilst the C IV emission 
line is also blue-shifted by $\sim5000$~km~s$^{-1}$. 
Thus PDS 456 may just reside at the extreme end of the BAL quasar 
population \citep{Turnshek88, Weymann91}.

Similar highly ionized X-ray outflows have recently 
been discovered in the non-BAL radio-quiet quasars 
PG 1211+143 and PG 0844+349 \citep{Pounds03a, Pounds03b}, 
suggesting that these properties are not unique  
to UV identified BAL quasars. We also note that strong iron K absorption has 
also been observed in the NLS1s
1H~0707-495 and IRAS~13324-3809 \citep{Boller02, Boller03} and also in 
some X-ray binaries, e.g. GRS 1915+105 \citep{Lee02}, GX~13+1 
\citep{Sidoli02}, X~1624-490 \citep{Parmar02}. 
In the quasars PDS 456, PG 1211+143 and PG 0844+349, the 
X-ray columns are near to $10^{24}$cm$^{-2}$, the outflow 
velocities measured are $\sim0.1c$, whilst the ionizations are 
also similar ($\xi\sim10^{3}$). 
Using PG~1211+143 as an example, \citet{King03} have recently 
demonstrated, for radiatively driven outflows, 
that objects accreting near to the 
Eddington rate are likely to contain massive 
($\dot{M}_{\rm out}\sim\dot{M}_{\rm edd}$) Compton-thick winds. 
Indeed when the outflow rate is of the same order as the accretion rate, 
the flow can become optically thick at $\sim10-100R_{g}$. 
This can account for both the high columns and velocities of the 
X-ray outflows, as well as provide a natural 
source of thermalized photons for the UV and soft X-ray excesses observed 
in these quasars. 

However other scenarios are possible, for instance \citet{R02} recently 
used a magnetic flare model to explain the rapid X-ray variability in 
PDS 456, through coherent flaring events occuring 
when the accretion rate is close to Eddington.
Indeed it is conceivable that the outflowing matter in PDS 456 is the 
ejecta associated with these flares.
Interestingly the output of the magnetic events 
($\sim10^{46}$~erg~s$^{-1}$) in PDS 456 is very similar to the kinetic 
power of the outflow, so this scenario seems energetically plausible. 
Overall the extreme properties of the outflow in PDS 456 seem linked 
to its high accretion rate \citep{R00}, which appears to 
drive the energetics of its unusually massive outflow. 

\section{Acknowledgments}

This paper is based on observations obtained with XMM-Newton, an ESA
science mission with instruments and contributions directly funded by
ESA Member States and the USA (NASA). We thank Ken Pounds and 
Andrew King for discussions and Kim Page for performing 
spectral fits. James Reeves thanks the Leverhulme Trust
for their support.

\clearpage

\begin{figure*}
\rotatebox{-90}{
\epsscale{0.7}
\plotone{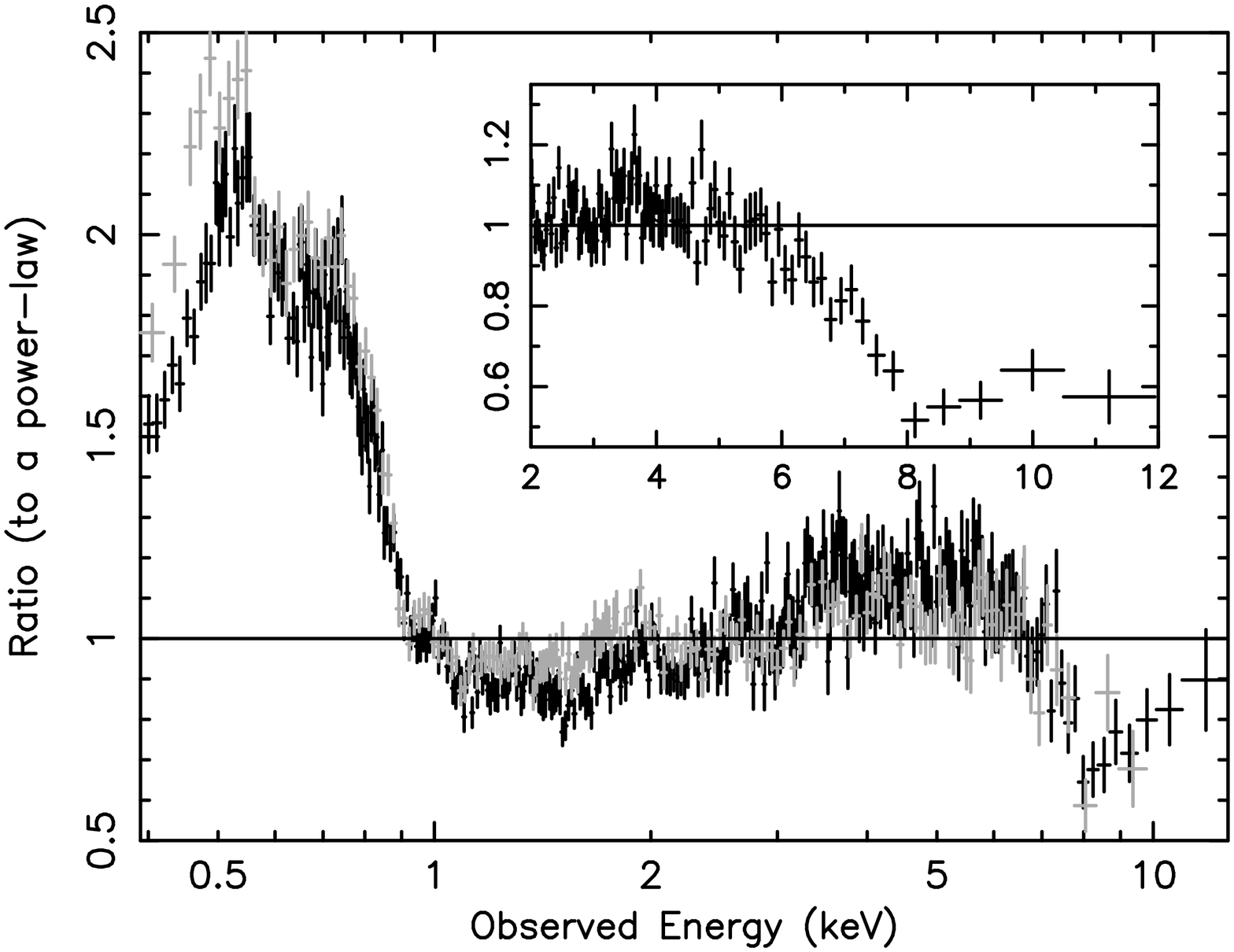}}
\caption{Ratio of the broad band X-ray spectrum of PDS 456 to a simple 
Galactic absorbed power-law model (with $\Gamma=2.0$). 
EPIC-pn data are plotted in black, EPIC-MOS in gray. 
A clear soft X-ray excess is observed 
below 1 keV, as well as a large decrease in counts above 7 keV. 
The inset plots the data/model ratio to a power-law fit to the 
2-12 keV EPIC-pn spectrum in greater detail. 
A 50\% drop in counts above the iron K-shell band is 
observed. Note the energy scale is plotted in keV in the observed frame.}
\end{figure*}

\begin{figure*}
\begin{center}
\rotatebox{-90}{\includegraphics[height=11.5cm]{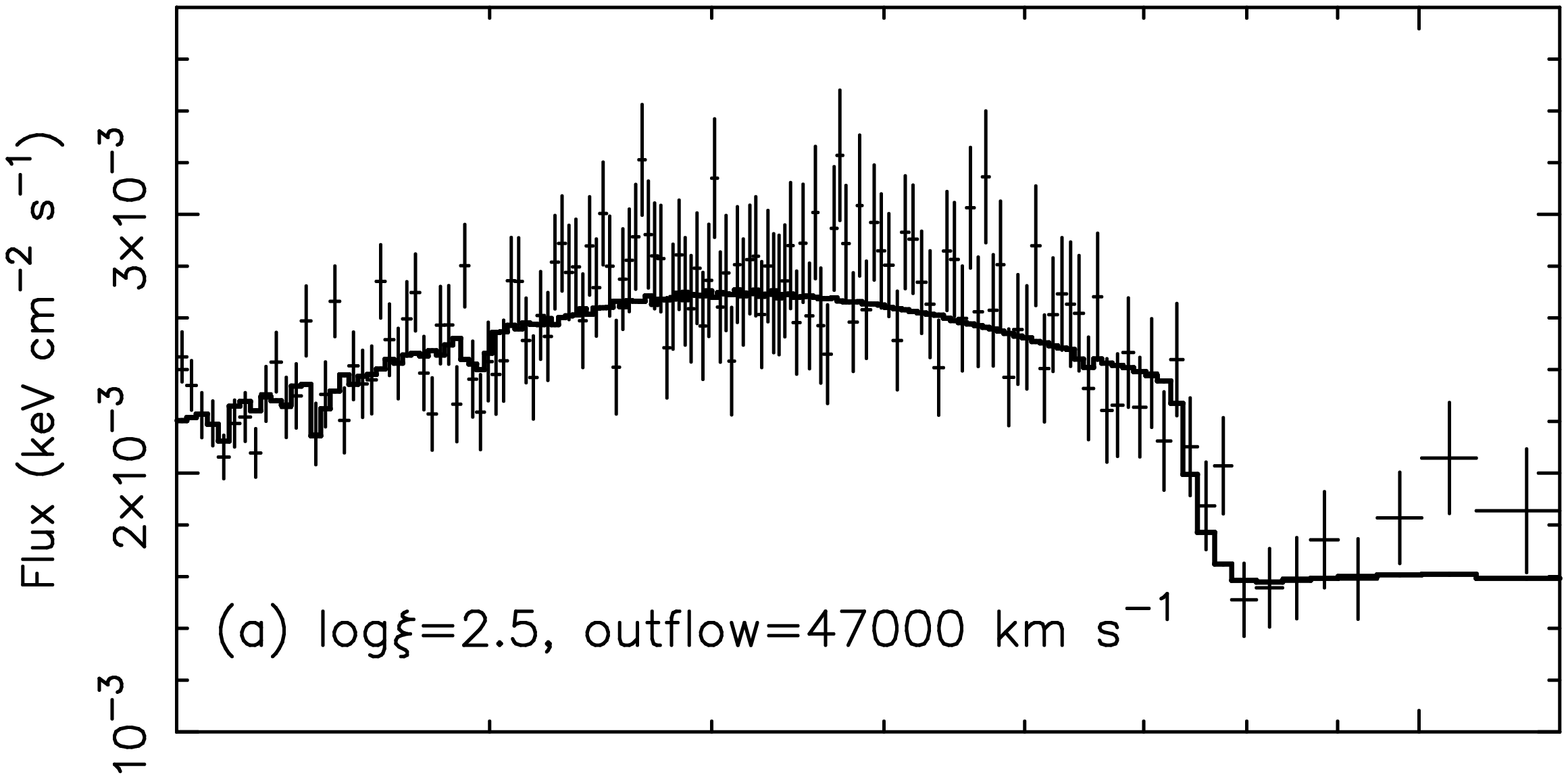}}
\rotatebox{-90}{\includegraphics[height=11.5cm]{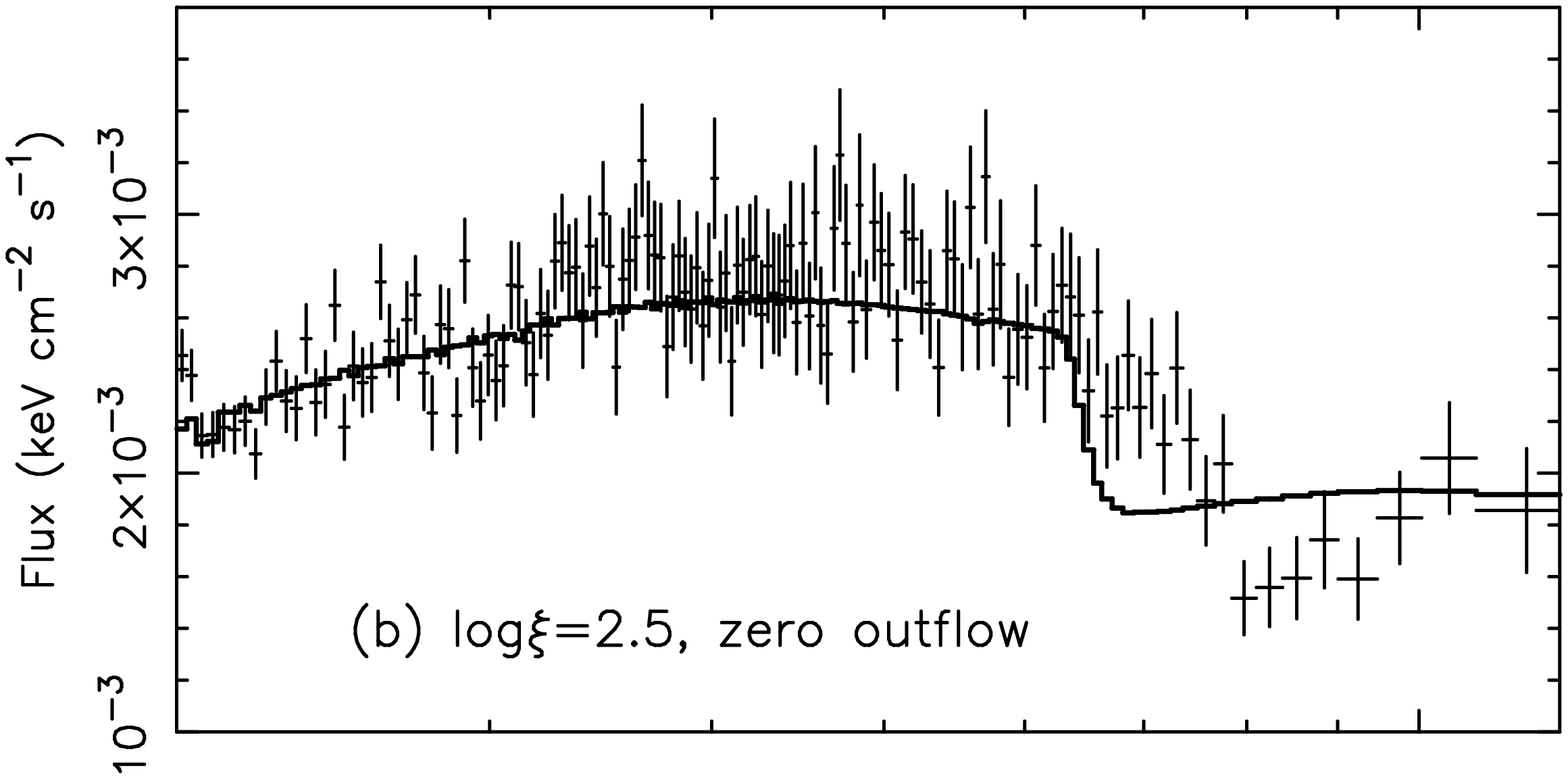}}
\rotatebox{-90}{\includegraphics[height=11.5cm]{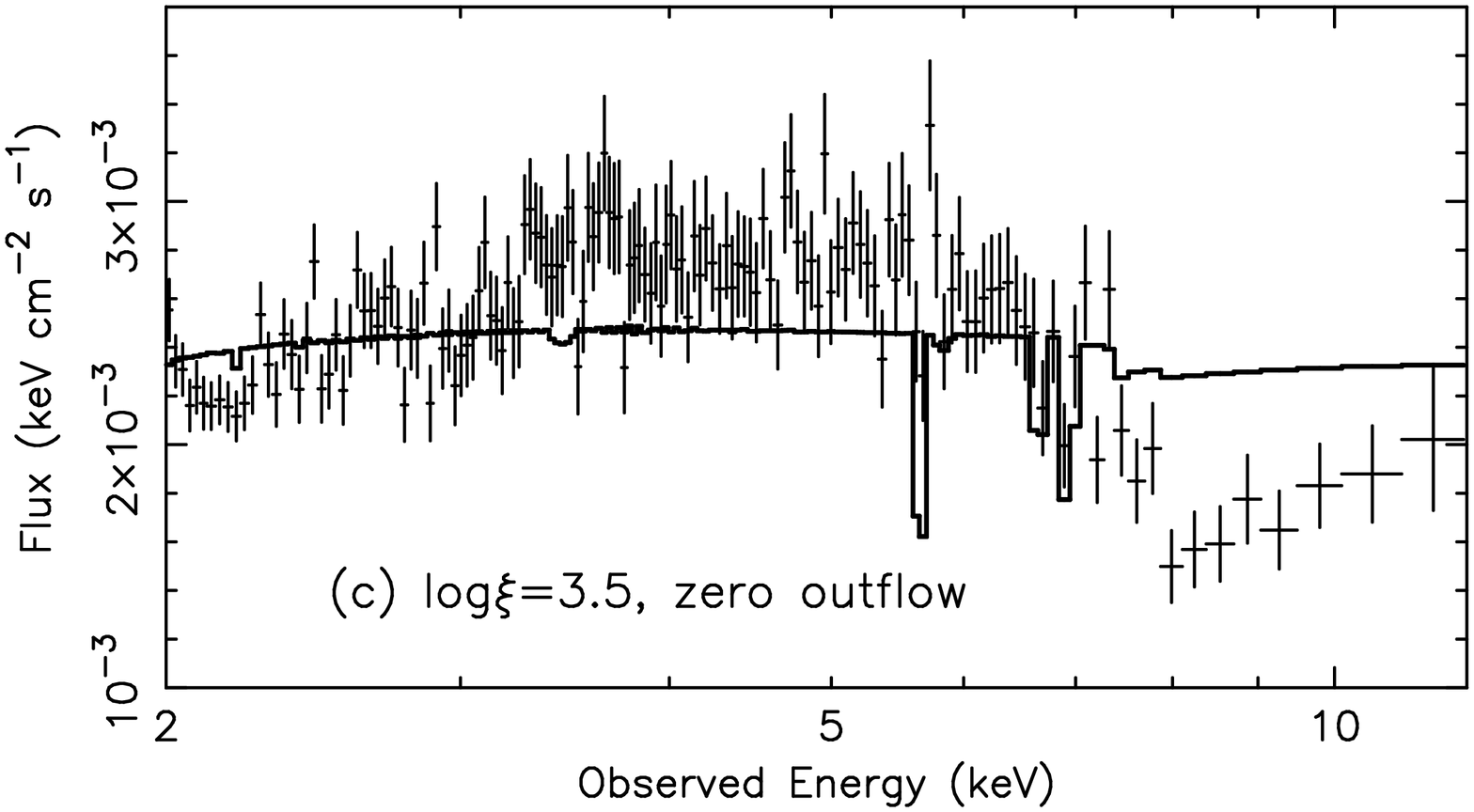}}
\end{center}
\caption{Warm absorber fits to the EPIC-pn data of PDS 456 using the 
\textsc{xstar} 
photoionization code. (a) shows the best fit model (solid line) fitted 
to the EPIC-pn data (crosses), which requires an 
outflow velocity of $\sim50000$~km~s$^{-1}$ and an ionization parameter of 
log~$\xi=2.5$; (b) shows the same model, but with no outflow velocity, 
resulting in a significantly worse fit; (c) the same model, but 
with no outflow 
and a higher ionization parameter of log~$\xi=3.5$. At this high ionization 
the model cannot reproduce the shape of the iron K-shell absorption seen in 
PDS 456.}
\end{figure*}

\begin{figure*}
\rotatebox{-90}{
\epsscale{0.7}
\plotone{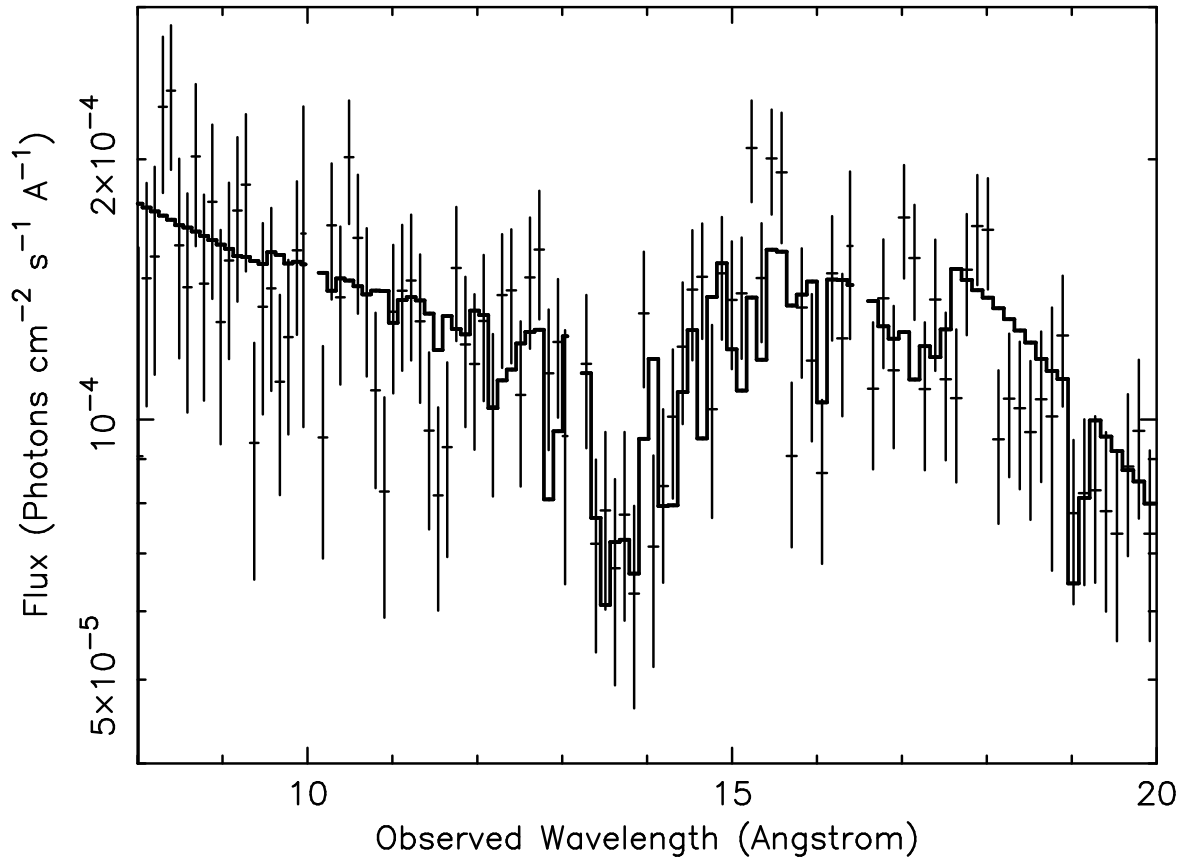}}
\caption{XMM-Newton RGS-2 spectrum of PDS 456, the data points are shown 
as crosses. 
A strong absorption trough is present at 12-15~\AA, observed frame. 
The best fit 
photoionization model (solid line) is superimposed on top of the data points, 
a high outflow velocity of $\sim50000$~km~s$^{-1}$ is required for 
the absorber.  
Note due to a malfunctioning chip, RGS-1 does not 
containing data in the 12-15~\AA\ range and thus is not plotted. 
Most of the opacity of the 
absorber results from L-shell transistions of iron in the ionization range 
Fe~\textsc{xvii} -- Fe~\textsc{xxiv}.}
\end{figure*}

\end{document}